\documentclass[10pt]{iopart}
\usepackage{iopams}  

\usepackage{graphicx}  
\usepackage{graphicx}  

\def\msun{\,{\rm M_\odot}}

\newcommand\beq{\begin{equation}}
\newcommand\eeq{\end{equation}}

\begin{document}

\title[Gravitational 
Waves from Coalescing Massive Black Holes]{{\it LISA} detection of massive black hole binaries: imprint of seed populations 
and of exterme recoils}





\author{A Sesana$^1$,
\ M Volonteri$^2$\ and \ F Haardt$^3$\
}

\address{$^1$\ Center for Gravitational Wave Physics at the Pennsylvania State University, University park, State College, PA 16802, USA}

\address{$^2$\ Department of Astronomy, University of Michigan, 500 Church Street, Ann Arbor, MI 48109, USA}

\address{$^3$\ Dipartimento di Fisica e Matematica, Universit\`a dell'Insubria,
via Valleggio 11, 22100 Como, Italy}

\begin{abstract}
All the physical processes involved in the formation, merging, and accretion history of 
massive black holes along the hierarchical build--up of cosmic structures are likely
to leave an imprint on the gravitational waves detectable by future
space--borne missions, such as {\it LISA}. We report here the results of 
recent studies, carried out by means of dedicated simulations of black hole build--up, 
aiming at understanding the impact on {\it LISA} observations of two ingredients that 
are crucial in every massive black hole formation scenario, 
namely: (i) the nature and abundance of the first black hole seeds and (ii)
the large gravitational recoils following the merger of highly
spinning black holes. We predict {\it LISA} detection rates spanning two 
order of magnitude, in the range 3-300 events per year, depending on the detail
of the assumed massive black hole seed model. On the other hand, large recoil velocities 
do not dramatically compromise the efficiency of {\it LISA} observations. The number of 
detections may drop substantially (by $\sim60\%$), in scenarios characterized by
abundant light seeds, but if seeds are already  massive and/or relatively rare, 
the detection rate is basically unaffected.   
\end{abstract}

\section{Introduction}
Massive black hole (MBH) binaries (MBHBs) are among the primary candidate 
sources of gravitational waves (GWs) at mHz frequencies  
~\cite{enelt, jaffe, uaiti, ses04, ses05}, the range probed by the space-based 
{\it Laser Interferometer Space Antenna} ({\it LISA}, ~\cite{bender}). 
Today, MBHs are ubiquitous in the nuclei of nearby galaxies ~\cite{mago}. If MBHs were 
also common in the past (as implied by the notion that many distant galaxies 
harbor active nuclei for a short period of their life), and if their host 
galaxies experience multiple mergers during their lifetime, as dictated by 
popular cold dark matter (CDM) hierarchical cosmologies, then MBHBs inevitably formed  
in large numbers during cosmic history. MBHBs that are able to coalesce in less than 
a then Hubble time give origin to the loudest GW signals in the Universe.
Provided MBHBs do not ``stall'', their GW driven inspiral will then 
follow the merger of galaxies and protogalactic structures at high redshifts. 
A low--frequency detector like {\it LISA} will be sensitive to GWs from coalescing binaries 
with total masses in the range $10^3-10^6\,\msun$ out to $z\sim 10-15$ ~\cite{hughes}. 

The formation and evolution of MBHs has been investigated recently by several 
groups in the framework of hierarchical clustering cosmology 
~\cite{MHN01, VHM03, KBD04}. {\it LISA} detection rate, ranging from a few to a few hundred per 
year, were derived in a number of papers ~\cite{uaiti, ses04, ses05, rw05}. 
A comprehensive understanding of the details of MBH formation and 
evolution are essential in assessing {\it LISA} detection efficiency
and in planning sensible data analysis strategies. 

In two recent papers ~\cite{ses07a, ses07b}, we considered in detail two important 
ingredients of the MBH formation route. (i) 
{\it The nature and abundance of the first black hole seeds}.
Our understanding of seed black hole formation is 
extremely poor. There are several proposed formation mechanisms resulting
in a broad spectrum of possible seed populations ~\cite{madres01, KBD04, bvr06, ln06}.
Following ~\cite{ses07a}, we investigate different physically and cosmologically 
motivated seed formation routes, showing their imprint on the
expected MBHB coalescence rate and hence on {\it LISA} detection. 
(ii) {\it Extreme gravitational recoils}. Recent relativistic numerical simulations of 
merging spinning black hole binaries have shown that the remnant may get a kick of 
the order of a few thousand km/s ~\cite{tm07, cetal07}, which is 
likely to eject it even from the center of a giant elliptical
(escape velocity $\gtrsim$ 2000 km/s), with important astrophysical
implications ~\cite{mad04, merr04, vol07}. We incorporate the effect of extreme gravitational 
recoils following the merger of highly spinning black holes in our hierarchical
models ~\cite{ses07b}, exploring its consequences on the GW detection side.

We highlight in this paper the main results reported in ~\cite{ses07a, ses07b}, 
focusing on their implications for {\it LISA} detections and assessing 
{\it LISA} capability to place constraints on MBH formation scenarios, looking 
for reliable diagnostics to discriminate between the different models. The plan
of the paper is as follow. In Section 2, we describe the general framework
of MBHB formation in hierarchical scenarios. We briefly discuss the issue of GW
detection with {\it LISA} in Section 3, and then in Section 4 and 5 we discuss
the impact of the seed black hole population and of the gravitational recoil prescription
on {\it LISA} observations. We summarize our findings in Section 6.    

\section{Hierarchical models of black hole formation}

In the hierarchical framework of structure formation ~\cite{wr78, pb82}, MBHs grow starting from 
pregalactic seed black holes formed at early times. MBH evolution then follows
the merging history of their host galaxies and dark matter halos. The merger process 
would inevitably form a large number of MBHBs during cosmic history.
In our models we apply the extended Press \& Schechter EPS formalism ~\cite{PS74, LC93, ST99} to the hierarchical assembly 
of dark matter halos, using a range of prescriptions for the evolution of the 
population of MBHs residing in the halo centers. The halo hierarchy 
is followed by means of Monte Carlo realizations
of the merger hierarchy. Each model is constructed tracing backwards the 
merger hierarchy of 220 dark matter halos in the mass range 
$10^{11}-10^{15}\msun$ up to $z=20$ ~\cite{VHM03}, 
then populating the halos with seed black holes and following their 
evolution to the present time. Nuclear activity is triggered by halo mergers: 
in each major merger the more massive hole accretes gas until its mass scales 
with the fifth power of the circular velocity of the host halo,  
normalized to reproduce the observed local correlation 
between MBH mass and velocity dispersion ($m_{\rm BH}-\sigma_*$ relation ~\cite{tr02}).  
Gas accretion onto the MBHs is assumed to occur at a fraction of the Eddington rate.
In the boundaries given by this general framework, there is a certain freedom in the choice 
of the seed masses, in the accretion prescription, and in the MBHB coalescence efficiency. 

\subsection{MBHB dynamics}
During a galactic merger, the central MBHs initially share their fate with the host galaxy. 
The merging is driven by dynamical friction, which has been shown to efficiently 
merge the galaxies and drive the MBHs in the central regions of the newly formed 
galaxy when the mass ratio of the satellite halo to the main halo is sufficiently 
large ~\cite{kz05}.  The efficiency of dynamical friction decays 
when the MBHs get close and form a binary.  In gas-poor systems, the subsequent 
evolution of the binary may be largely determined by three-body interactions 
with background stars
~\cite{bbr80}, leading to a long coalescence timescale.  
In gas rich high redshift halos, the orbital evolution of the central MBH is 
likely dominated by dynamical friction against the surrounding gaseous medium. 
The available simulations ~\cite{E04, D06} show that the binary may shrink to 
about parsec or slightly subparsec scale by dynamical friction against the gas, 
depending on the gas thermodynamics. We have assumed here that, if a hard MBH 
binary is surrounded by an accretion disc, it coalesces  instantaneously owing 
to interaction with the gas disc. If instead there is no gas readily available, 
the binary will be losing orbital energy to the stars, using the scheme 
described in ~\cite{VHM03}.

Assuming efficient coalescence for the MBH pairs, for each of the 220 halos 
all the coalescence events happening during the cosmic history 
are collected. The outputs are then weighted 
using the EPS halo mass function and integrated over the observable 
volume shell at every redshift to obtain numerically the coalescence 
rate of MBHBs as a function of black hole masses and redshift. 

\section{Gravitational wave signal}
Full discussion of the GW signal produced by an inspiraling 
MBHB can be found in ~\cite{ses05}, along with all the relevant references.
Here we just recall that a MBHB at (comoving) distance $r(z)$
with chirp mass ${\cal M}=m_1^{3/5}m_2^{3/5}/(m_1+m_2)^{1/5}$ ($m_1>m_2$ are the 
two MBH masses) generates a GW signal with a characteristic strain given by:
\begin{equation}
h_c=\frac{1}{3^{1/2}\pi^{2/3}}\,\frac{G^{5/6}
{{\cal M}}^{5/6}}{c^{3/2} r(z)}\,f_r^{-1/6},
\label{eq1h_c}
\end{equation}
where $G$ and $c$ have their standard meaning.
An inspiraling binary is then detected if the signal-to-noise ratio ($S/N$)
{\it integrated over the observation} is larger than a given 
detection threshold, where the optimal $S/N$ is given by ~\cite{FH98}
\begin{equation}
S/N=\sqrt{ \int d\ln f \, \left[
\frac{h_c(f_r)}{h_{\rm rms}(f)} \right]^2}. 
\label{eqSN}
\end{equation}
Here, $f=f_r/(1+z)$ is the (observed) frequency emitted at time 
$t=0$ of the observation, and the integral is performed over the
frequency interval spanned by the shifting binary during the observational time. 
Finally, $h_{\rm rms}=\sqrt{5fS_h(f)}$ is the effective rms noise of the 
instrument; $S_h(f)$ is the one-sided noise spectral density, and the
factor $\sqrt{5}$ takes into account for the random directions and 
orientation of the wave; $h_{\rm rms}$ is obtained by adding in the instrumental 
noise contribution (given by e.g. the Larson's online sensitivity curve generator 
http://www.srl.caltech.edu/$\sim$shane/sensitivity), and the confusion noise from  
unresolved galactic ~\cite{N01} and extragalactic 
~\cite{FP03} WD--WD binaries. Notice that extreme mass-ratio inspirals 
(EMRIs) could also contribute to the confusion noise in the mHz frequency range ~\cite{BC04}. 
All the results shown in the following sections assume a {\it LISA} operation time of 
3 years, a cut-off at $10^{-4}$ Hz in the instrumental sensitivity and a detection 
integrated threshold of $S/N=5$ (equation~\ref{eqSN}).

\section{Seed black hole population imprint}

In this section we investigate the influence of different seed population prescriptions
on {\it LISA} detections. In all the models described below, we assume non spinning 
binaries; the coalescence remnant thus experiences only a mild recoil, that is never larger 
than $\sim$ 250 km/s ~\cite{fhh04, baker06}. The effects of extreme gravitational recoil 
associated to the coalescence of highly spinning MBHs will be separately explored
in Section 5.

\subsection{Description of the models}
Several scenarios has been proposed for the seed black hole formation:
seeds of $m_{\rm seed}\sim $few$\times100\msun$ can form as remnants of metal free (PopIII) 
stars at redshift $\gtrsim20$ (as assumed by Volonteri, Haardt and Madau ~\cite{VHM03}, 
hereinafter VHM model), while intermediate--mass seeds ($m_{\rm seed}\sim10^5\msun$) can be the 
endproduct of the dynamical instabilities arising in  massive gaseous protogalactic 
disks in the redshift range $10\lesssim z \lesssim15$ (as investigated
by Koushiappas, Bullock and Dekel ~\cite{KBD04}, hereinafter KBD model; 
or by Begelman, Volonteri and Rees ~\cite{bvr06}, hereinafter BVR model).

In the VHM model, pregalactic seed holes form with intermediate 
masses ($m_{\rm seed}\sim150\,\msun$) as remnants of the first generation of massive 
metal-free stars with $m_*>260\,\msun$ that do not disappear as 
pair-instability supernovae ~\cite{madres01}. We place them in isolation 
within halos above $M_H=1.6\times 10^7\,\msun$ collapsing 
at $z=20$ from rare $>3.5\sigma$ peaks of the primordial density 
field. While $Z=0$ stars with $40<m_*<140\,\msun$ are also predicted 
to collapse to MBHs with masses exceeding half of the initial stellar mass 
~\cite{hw02}, the merger rate of MBHBs in the mass range relevant to {\it LISA}
observations is not very sensitive to the precise choice for the seed hole mass.

A different class of models assumes that MBH seeds form already massive. 
In the KBD model, seed MBHs form from the low angular momentum tail of material in halos 
with efficient molecular hydrogen gas cooling. MBHs with mass
\beq 
m_{\rm seed}\simeq5\times10^4\msun(M_H/10^7\msun)(1+z/18)^{3/2}(\lambda/0.04)^{3/2}
\label{KBD:mbh}
\eeq
form in in dark matter halos with mass  
\beq
M_H \gtrsim 10^7\msun (1+z/18)^{-3/2}(\lambda/0.04)^{-3/2}.
\eeq
We have fixed the free parameters in equation \ref{KBD:mbh} by requiring an acceptable match
with the luminosity function of quasars at $z<6$. 
Here $\lambda$ is the so called spin parameter, which is a measure of the angular 
momentum of a dark matter halo $\lambda \equiv J |E|^{1/2}/G M_H^{5/2}$, where 
$J$, $E$ and $M_H$ are the  total angular momentum, energy and mass of the halo.  
The angular momentum of galaxies is believed to have been acquired by tidal 
torques due to interactions with neighboring halos. The distribution of spin 
parameters found in numerical simulations is well fit by a lognormal 
distribution in $\lambda_{\rm spin}$, with mean  $\bar \lambda_{\rm spin}=0.04$ 
and standard deviation $\sigma_\lambda=0.5$ ~\cite{B01, vdB02}. 
We have assumed that the MBH formation process proceeds 
until $z\approx15$.

In the BVR model, black hole  seeds form in halos  subject to runaway gravitational instabilities, 
via the so called "bars within bars" mechanism ~\cite{SFB89}. 
We assumed here, as in BVR, that runaway instabilities are efficient 
only in metal free halos with virial temperatures $T_{\rm vir} \gtrsim 10^4$K. 
The "bars within bars"  process produces in the center of the halo  a "quasistar" 
(QSS) with a very low specific entropy.  When the QSS  core  collapses, it leads 
to a seed black hole  of a few tens solar masses.  Accretion from the QSS 
envelope surrounding the collapsed core can however build up a substantial 
black hole mass very rapidly  until it reaches a mass of the order the "quasistar"  
itself, $m_{\rm QSS}\simeq 10^4-10^5 \msun$.   The seed black hole accretion rate adjusts 
so that the feedback energy flux equals the Eddington limit for the quasistar 
mass; thus, the black hole grows at a super-Eddington rate as long as 
$m_{QSS} > m_{\rm seed}$. The result is that $m_{\rm seed} (t) \sim  4 \times 10^5  (t/10^7 \ {\rm yr})^2  
M_\odot$ i.e., $m_{\rm seed} \propto t^2$. In metal rich halos star formation 
becomes efficient, and depletes the gas inflow before the conditions for QSS 
(and MBH) formation are reached. BVR 
envisage that the process of MBH formation stops when gas is sufficiently 
metal enriched. We consider here two scenarios, one in which star formation exerts a high 
level of feedback and ensures a rapid metal enrichment (BVRhf), one in 
which feedback is milder and halos remain metal free for longer (BVRlf).
In the former case MBH formation ceases at $z\approx 18$, in the latter at $z\approx 15$.
The BVRhf model appears to produce barely enough MBHs to reproduce the 
observational constraints (ubiquity of MBHs in the local Universe, luminosity function of quasars). 
We consider it a very strong lower limit to the number of seeds that need to be formed 
in order to fit the observational constraints. 

   \begin{figure}
   \centering
   \resizebox{\hsize}{!}{\includegraphics[clip=true]{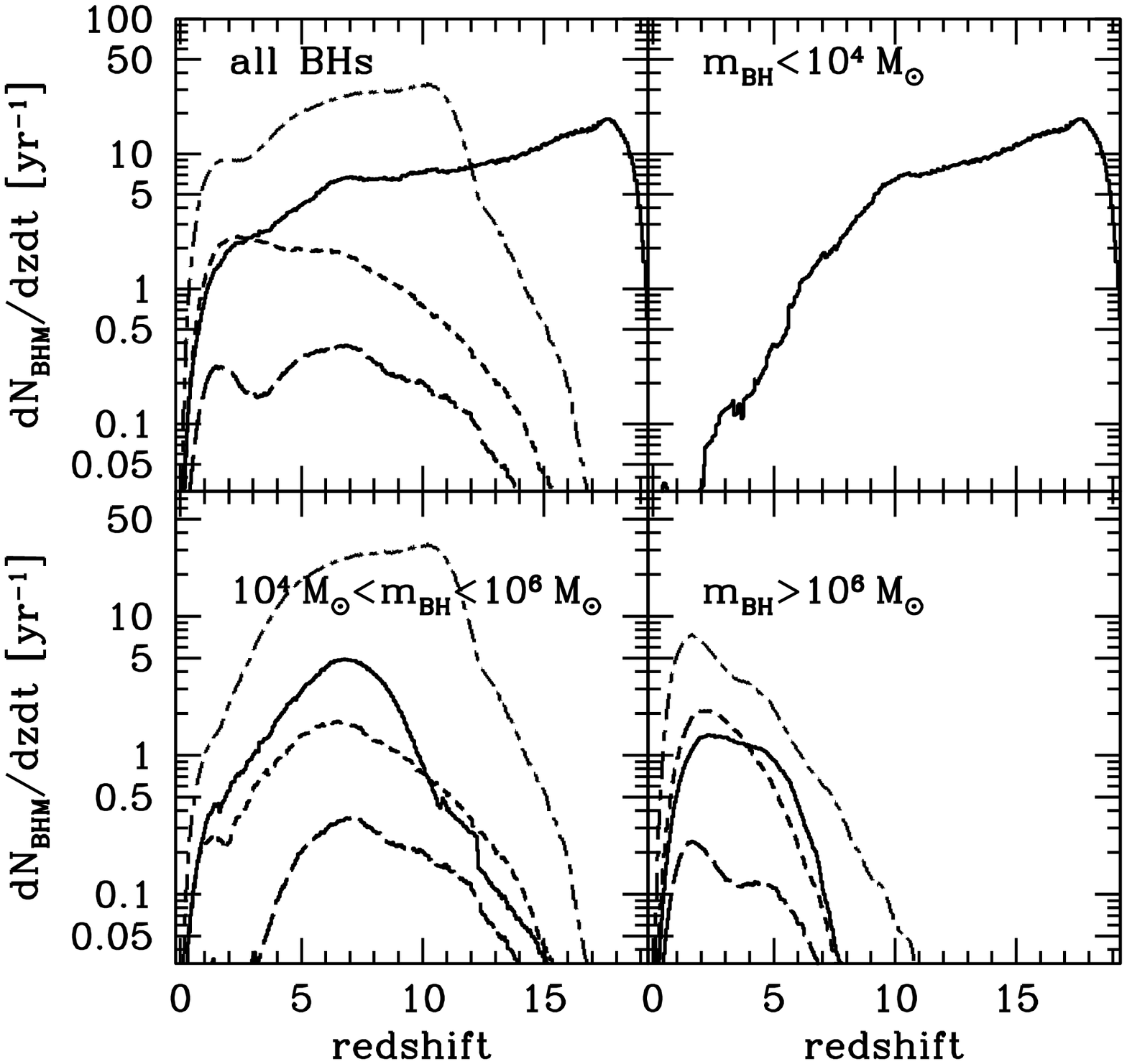}
   \includegraphics[clip=true]{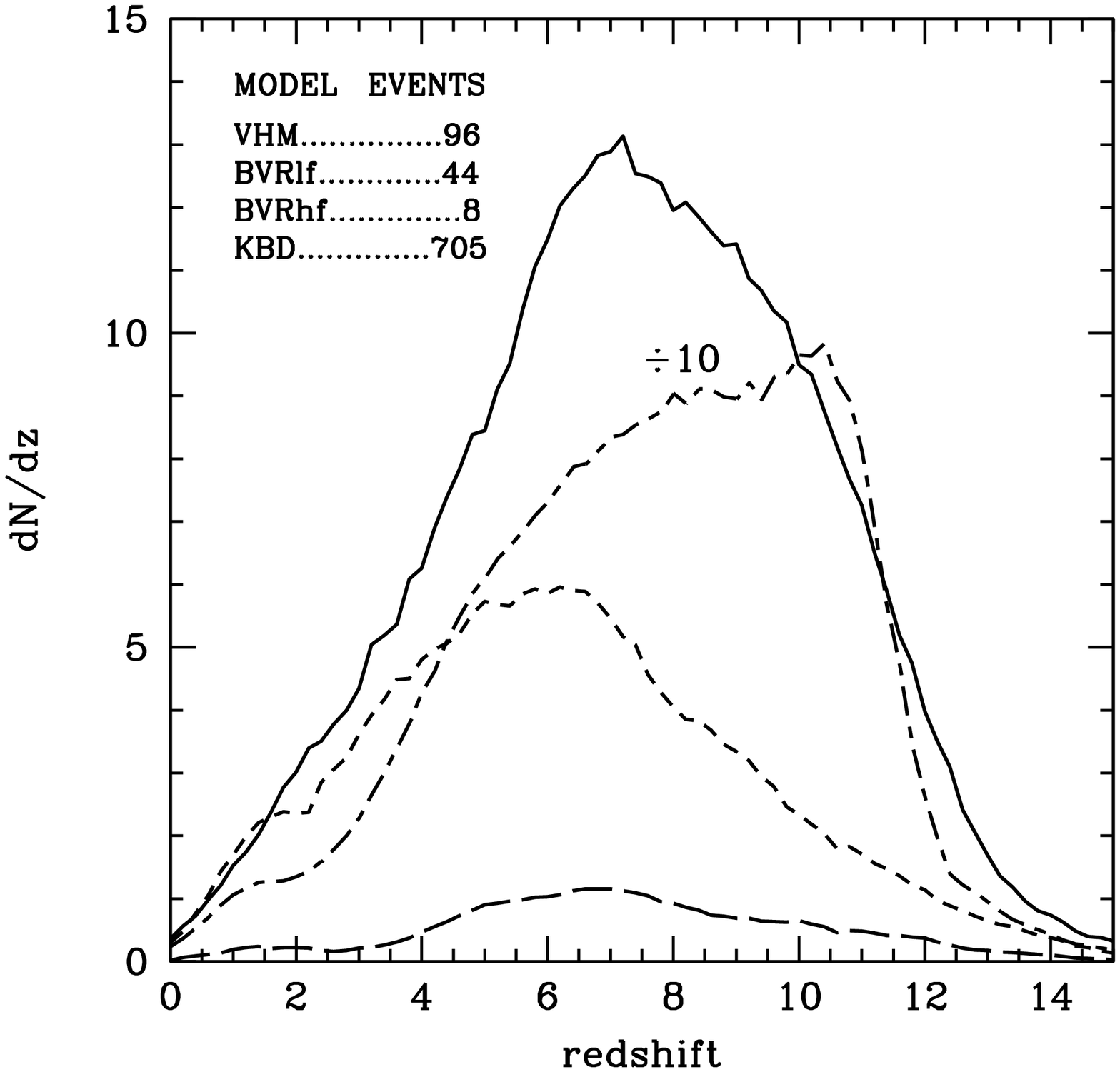}}
     \caption{{\it Left panel}: number of MBHB coalescences per observed year at $z=0$, per unit redshift, 
in different $m_{\rm BH}=m_1+m_2$ mass intervals. {\it Solid lines}: VHM model;
{\it short--long dashed lines}: KBD model; {\it short--dashed lines}: 
BVRlf model; {\it long--dashed lines}: BVRhf model.
{\it Right panel}: redshift distribution of MBHBs resolved with 
$S/N>5$ by {\it LISA} in a 3-year mission. Line style as in the left panel.
The number of events predicted by KBD model
({\it long--short dashed curve}) is divided by a factor of $10$.
The top-left corner label lists the total number of expected detections.
}  
        \label{figseed}
    \end{figure}

\subsection{Coalescence rates and {\it LISA} detections}

Different seed populations would inevitably leave a peculiar imprint on
the merger rate of MBHBs relevant to {\it LISA}.
Left panel of figure \ref{figseed} shows the number of MBH binary coalescences per unit 
redshift per unit {\it observed} year, $dN/dzdt$, predicted by the five models we tested. 
Each panel shows the rates for 
different $m_{\rm BH}=m_1+m_2$ mass intervals. The total coalescence rate
spans almost two orders of magnitude
ranging from $\sim 3$ yr$^{-1}$ (BVRhf) to $\sim 250$ yr$^{-1}$ (KBD).
As a general trend, coalescences of more massive MBHBs peak at lower 
redshifts (for all the models the coalescence peak in the case 
$m_{\rm BH}>10^6\msun$ is at $z\sim2$). 
Note that there are no merging MBHBs with $m_{\rm BH}<10^4\msun$ 
in the KBD and BVR models.

   \begin{figure}
   \centering
   \includegraphics[width=3.5in]{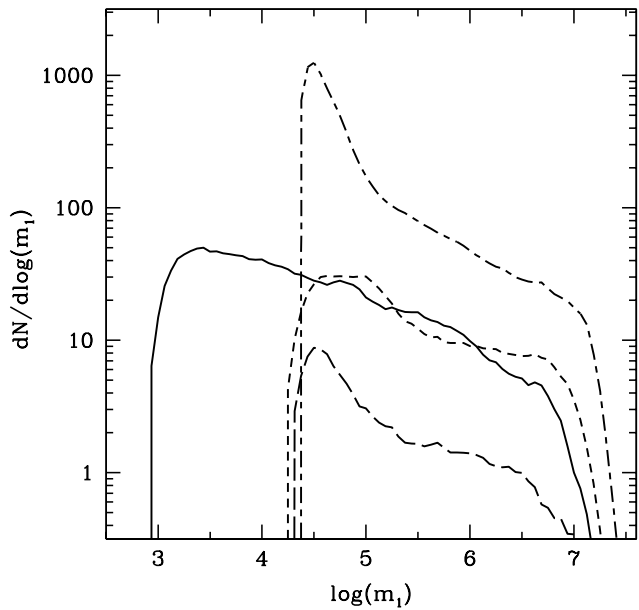}
      \caption{Mass function of the more massive member 
of MBHBs resolved with $S/N>5$ by {\it LISA} in a 3-year mission. 
Line style as in figure \ref{figseed}. All curves are normalized such as the integral 
in $d\log(m_1)$ gives the number of detected events.
             }
         \label{m1dist}
   \end{figure}

Right panel of figure~\ref{figseed} shows the redshift distribution of {\it LISA} MBHB 
detections. The KBD model results in a number of events ($\simeq 700$) that is more than an 
order of magnitude higher than that predicted by other models, 
with a skewed distribution peaked at sensibly high redshift, $z \gtrsim 10$. 
It is interesting to compare the {\it number of detections} with the {\it total number 
of binary coalescences}  predicted by the different formation models.
The KBD model produces $\simeq 750$ coalescences, the VHM model $\simeq 250$, 
and the two BVR models just few tens. A difference of a factor $\simeq 3$  
between the KBD and the VHM models in the total number of coalescences,
results in a difference of a  factor of $\simeq 10$ in the {\it LISA} detections, 
due to the different mass of the seed black holes. 
Almost all the KBD coalescences involve massive binaries ($m_1 \gtrsim 10^4 \msun$), 
which are observable by  {\it LISA}. The  KBD  and BVR models differ for 
the sheer number of MBHs. The halo mass threshold in the KBD model is well below 
(about 3 orders of magnitude) the BVR one, the latter requiring halos with virial 
temperature above $10^4$K. In a broader context,  results pertaining to the 
KBD model describe the behavior  of families of models where efficient 
MBH formation can happen also in mini-halos where the source of cooling is molecular hydrogen. 

It is difficult, on the basis of the redshift distributions of detected binaries only, 
to discriminate between heavy and light MBH seed scenarios.  
Although the VHM and BVRlf models predict a different number of observable sources,
the uncertainties in the models are so high, that a difference 
of a factor of two (96 for the VHM model, 44 for the BVRlf model) cannot be 
considered a safe discriminant. Moreover the redshift distributions are quite
similar, peaked at $z\simeq6-7$ and without any particular feature in the shape.
In ~\cite{ses05} we showed that {\it LISA} will be sensitive to 
binaries with masses $\lesssim10^3\msun$ up to redshift ten. Hence the 
discrimination between heavy and light MBH seed scenarios should be 
easy on the basis of the mass function of detected binaries.
This is shown in figure~\ref{m1dist}. As expected, in the VHM model, the mass 
distribution extends to masses $\lesssim10^3\msun$, giving a clear and unambiguous 
signature of a light seed scenario. VHM  predict that  many detections (about 50$\%$) involve 
low mass binaries ($m_{\rm BH}<10^{4}\msun$) at high redshift ($z>8$). These sources
are observable during the inspiral phase, but their frequency at the last stable orbit $f_{\rm LSO}$ is too high for 
{\it LISA} detection (see ~\cite{ses05}, figure 2). 
Heavy seed scenarios predict instead that the GW emission at $f_{\rm LSO}$, and the subsequent
plunge are always observable for all binaries. 

\section{Gravitational recoil imprint}

In all the models detailed in the previous section, we implemented a 'conservative' recoil prescription
appropriate to mergers of non spinning black holes. In the following, we quantify, for 
selected seed scenarios, the maximum effect that gravitational recoil may have on {\it LISA} 
detection rates.

\subsection{Description of the models}

We focus here on two specific models described in Section 4.1,
that are representative of these two classes of MBH seed formation scenarios: 
the VHM and the BVRlf models. For both of them, we consider two cases that bound the effect 
of recoil in the assembly of MBHs and, as a consequence, LISA events: (i) no gravitational 
recoil takes place and (ii) maximal gravitational recoil is associated to every MBHB merger, 
using the model by Volonteri 2007 ~\cite{vol07}, which is based on the estimates reported by Campanelli 
at al. 2007 ~\cite{cetal07}. For the latter we use the merger tree realizations 
presented in ~\cite{vol07}. The model takes into account consistently for the 
cosmic evolution of the mass ratio distribution of merging binaries and 
of their spin parameters (see discussion in ~\cite{vol07}). 
The spin orientations during each merger are instead always in the 
configuration leading to the maximum recoil according to  ~\cite{cetal07},
i.e., MBH spins are assumed to lie in the binary orbital 
plane counter-aligned one to each other. The recoil velocity is then 
computed according to equation 1 of  ~\cite{cetal07}  assuming 
$\Theta=\Theta_0$ (i.e., the maximum possible recoil velocity). 
We would like to emphasize that the prescription that we have chosen for (ii), 
and whose main features we have just summarized is the least favorable for 
GW observations and (probably) unlikely to occur in these 
extreme circumstances during MBH assembly. 

\subsection{Merger rates and {\it LISA} detections}

\begin{figure}
   \centering
   \resizebox{\hsize}{!}{\includegraphics[clip=true]{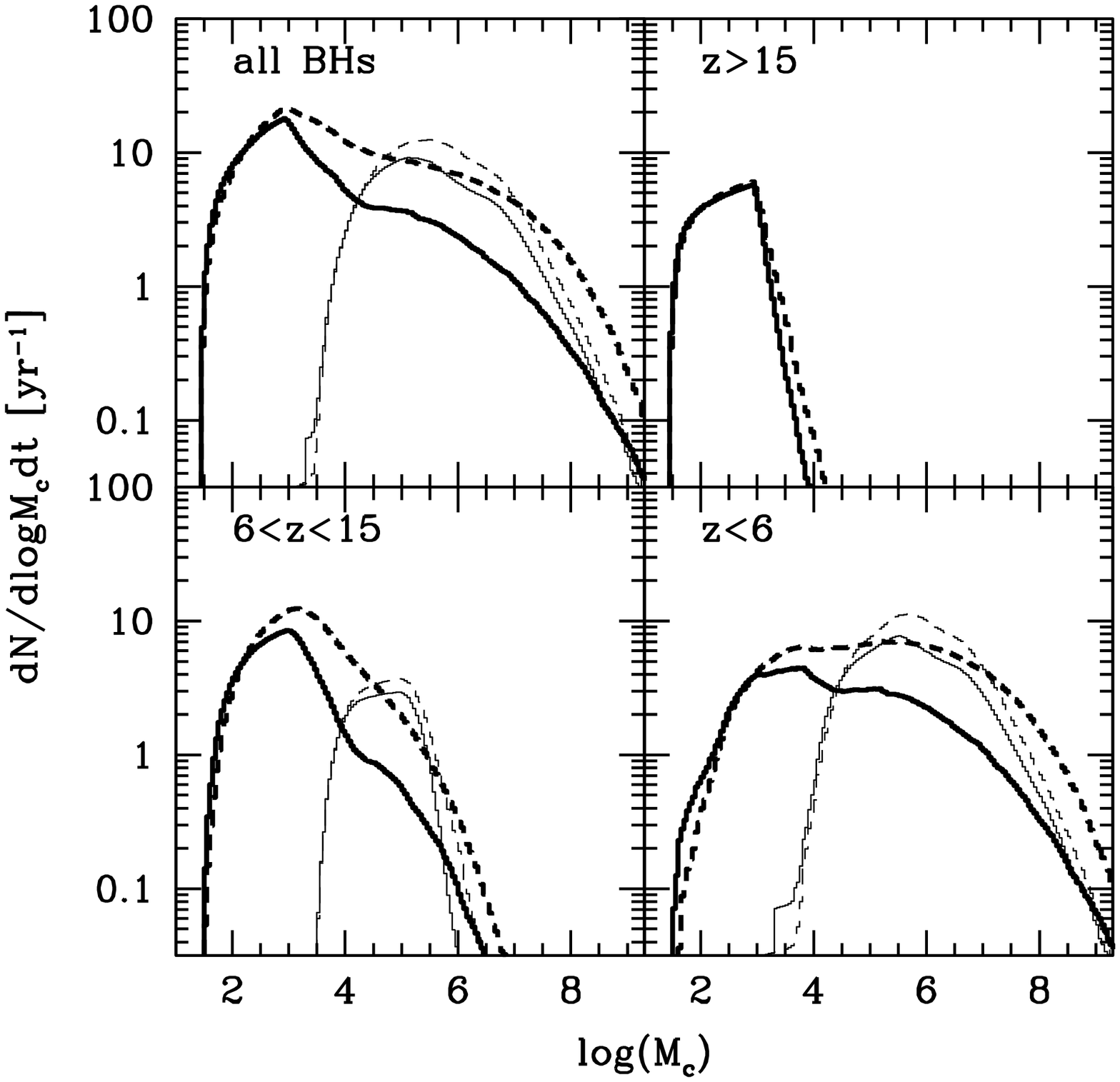}
   \includegraphics[clip=true]{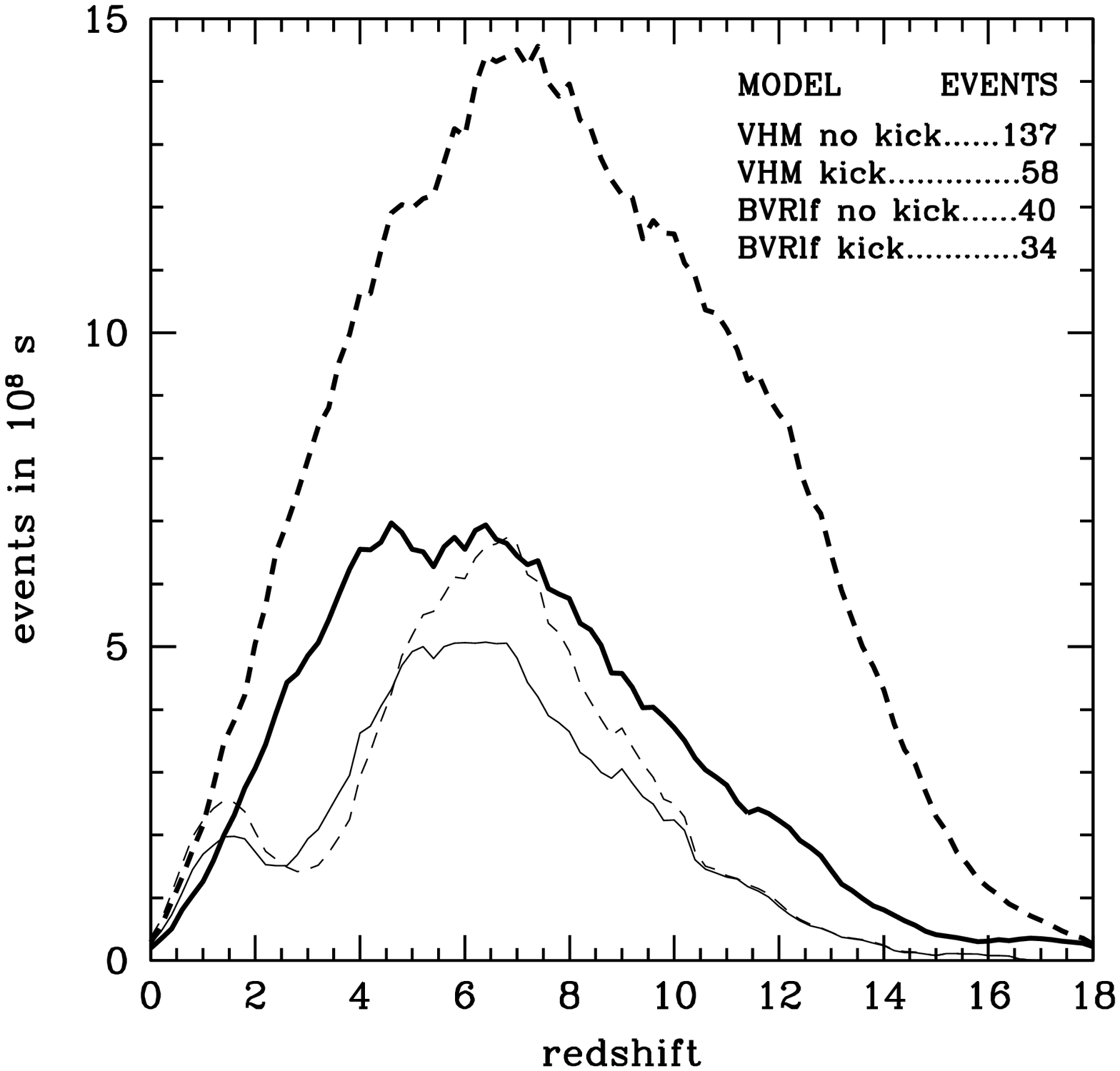}}
     \caption{{\it Left panel}: number of MBHB coalescences per observed year at $z=0$, per unit 
log chirp mass, 
in different redshift intervals. {\it Dashed lines}: GW recoil 
neglected; {\it solid lines}: extreme GW recoil included.
{\it Thick lines}: VHM model; {\it thin lines}: BVRlf model.
{\it Right  panel}: redshift distribution of MBHBs resolved with 
$S/N>5$ by {\it LISA} in a 3-year mission. Line style as in the left panel.
The top-right corner label lists the total number of expected detections.
}  
        \label{figrec}
    \end{figure}

Left panel of figure \ref{figrec} shows the number of MBH binary coalescences per unit 
${\rm log}{\cal M}$ per unit {\it observed} year, $dN/d{\rm log}{\cal M} dt$, 
predicted by the two models that we have considered, for both cases where
recoil is neglected and extreme recoil is taken into account. Each panel shows 
the rates for different redshift intervals. 
Note that when extreme recoil in included, the rate predicted by the BVRlf model at any 
redshift is only marginally affected, while the VHM model is more sensitive to the GW recoil:
at $z>15$, GW kicks do not affect the coalescence rate; 
on the contrary, at $z<15$, the rate drops by a factor of 
$\sim 3$ for ${\cal M}\gtrsim 10^3\msun$, 
if extreme kicks are included in the evolution. This is related to the fraction
of seeds that experience multiple coalescences during the MBH assembly
history. We can schematically think of the assembly history as a 
sequence of coalescence rounds. After each round 
extreme recoil depletes a large fractions of remnants, and the relative
importance of each subsequent round drops accordingly.
In the VHM model, about 65\% of the 
remnants of the first round will undergo a second round of coalescences,
so the second round has an important relative weight in the computation of the
total rate. When extreme recoil is taken into account, a large fraction of 
the first round remnants is ejected from their hosting halos. We
find that the effective fraction of remnants that can experience
a second coalescence drops to $\sim30\%$. This is the reason why the number
of coalescences involving light black holes (${\cal M}<10^3 \msun$) does not
drop at any redshift, while the number of coalescences involving more
massive binaries drops by a factor $\approx 3$. In the BVRlf scenario
seeds are rarer, and the fraction of first coalescence remnants that
participate to the second round is around 25\%; switching on
the extreme recoil has a significantly smaller impact on the global rate in this case.
Moreover, in this model seeds are more massive and the bulk of 
merging events happens at lower redshift, where the hosting halo 
potential wells are deeper and consequently larger kicks are needed
to eject the coalescence remnants. 
As a matter of fact, the seed abundance sets the mean number of major 
mergers that a seed is expected to undergo during the cosmic history, 
and this basically sets the ability of extreme kicks to reduce 
the coalescence rate. 

Right panel of figure ~\ref{figrec} shows the redshift distribution of MBHBs detected by 
{\it LISA}. The effect of extreme GW recoils on the source number 
counts drastically depends on the abundance and nature of the seeds. In the VHM 
model, the number of detectable sources drops by a factor $\sim 60\%$, and
the number of the potential {\it LISA} detections is reduced from
$\approx 140$, if the recoil is neglected, to $\approx 60$, if extreme 
recoil is included. Vice versa, the detection rate predicted by the BVRlf model
is only weakly affected by the extreme recoil prescription, and it
drops by about $15\%$ (from 40 to 34 events in 3 years of observation \footnote{Note that both numbers
are smaller than 44, which is the numbers of event found in Section 4 assuming a non-spinning recoil prescription. 
This is consistent with a Poissonian variance in the number of coalescences for different Monte Carlo 
realizations of the seed populations. The 15\% decrease found here (from 40 to 34) is instead 
computed starting from the same Monte Carlo realization of the seed population, 
and then applying different recoil recipes.} ). 
Note that though the overall number of coalescences in the VHM
model decreases only by about $25\%$ when extreme recoil is considered,
the number of {\it LISA} detections is reduced by a much larger
factor. This is because if the seeds are light, {\it LISA} can
not detect the bulk of the first coalescences of light
binaries happening at high redshift, that are responsible
for the major contribution to the coalescence rate and
are not affected by the recoil. {\it LISA}
can observe later events, involving more massive binaries,
that are largely suppressed by the MBH depopulation 
due to extreme GW kicks. In the BVRlf
model, on the other hand, seeds are more massive, and the second
coalescence round is less important; in this case, the {\it LISA} sensitivity
is sufficient to observe almost all the first coalescences, and
the number of detections is only mildly reduced. 

\section{Summary and conclusions}
Using dedicated Monte Carlo simulations of the hierarchical assembly 
of dark matter halos along the cosmic history, we have 
computed the expected gravitational wave signal from the evolving 
population of massive black hole binaries. 
We investigated the imprint of different seed black hole formation routes
and on extreme recoils on {\it LISA} detections. 

We found that a large fraction (depending on models) of coalescences 
will be directly observable by {\it LISA}, and on the basis of the detection 
rate, constraints can be put on the MBH formation process. 
Detection of several hundreds events in 3 years will be the sign of a 
efficient formation of heavy MBH seeds in a large fraction of high redshift halos (KBD).
On the other extreme, a low event rate, about  few tens in 3 years,  
is peculiar of scenarios where either the seeds are light, and many 
coalescences do not fall into the {\it LISA} band, or seeds are massive, 
but rare, as envisioned by, e.g., BVR (see also ~\cite{ln06}). 
In this case a decisive diagnostic is provided by the mass distribution 
of detected events. In the light seed scenario, the mass distribution
of observed binaries extend to $\sim 10^3\msun$, while there are no
sources with mass below $10^4\msun$ in the high seed scenario.  

We have then considered two specific MBH assembly models (VHM, BVRlf), representative
of two different MBH seed formation scenarios.  For both of them, we investigated two cases 
that bound the effect of recoil in the assembly of MBHs and, as a consequence, LISA events: 
(i) no gravitational  recoil takes place and (ii) maximal gravitational recoil is 
associated to every MBHB merger, using the model described in ~\cite{vol07}.
Our results show that at time it is not clear if {\it LISA} will be 
able to shed light on the importance of recoil in MBH assembly, 
even in this extreme case, since the uncertainty introduced in the 
number counts is at most of a factor of $\sim 3$, comparable with uncertainties
due to our ignorance in the MBH accretion history and in the
detailed dynamics of MBHBs (see, e.g., discussion in ~\cite{ses07a}).
On the other hand, this fact confirm that MBHBs are {\it LISA} safe targets; since 
extreme recoil effects increase with the seed abundance, we expect the drop in 
the detections to be more significant for those scenarios that predict a larger 
number of sources. 

In conclusions, from the point of detection of low frequency 
gravitational waves, massive black hole binaries are certainly one of 
the major target for the {\it LISA} mission, independently on the actual
seed black hole formation route and on the magnitude of the typical recoil
suffered by the remnants of binary coalecences. On the astrophysical ground, 
{\it LISA} will be a unique probe of the formation, accretion and merger of 
MBHs along the {\it entire} cosmic history of galactic structures.

\end{document}